\documentclass{desyproc}

\def\bb {\begin {eqnarray}}
\def\ee {\end {eqnarray}}

\usepackage{bm}

\def \ms {{\overline{\mbox{MS}}}}

\begin{document}
\title{Low-$x$
evolution of parton densities 
}
\author{{\slshape A.Yu. Illarionov$^1$, 
A.V. Kotikov$^2$\footnote{Speaker}}\\[1ex]
     $^1$Integrated Systems Laboratory, ETH,
CH-8092 Z\"urich, Switzerland\\
     $^2$Bogoliubov Laboratory of Theoretical Physics, JINR,
141980 Dubna, Russia}


\contribID{ZZ}
\confID{UU}
\desyproc{DESY-PROC-2012-YY}
\acronym{MPI@LHC 2011}
\doi
\maketitle

\begin{abstract}
It is shown that a Bessel-like
behaviour of 
the structure function $F_2$ at small $x$,
obtained for
a flat initial condition in the DGLAP evolution equations,
leads to 
good agreement with the 
deep inelastic 
scattering experimental data from HERA.

\end{abstract}


\section{Introduction}
The fairly
reasonable agreement between HERA data  
\cite{H1ZEUS}-\cite{:2009wt}
and the next-to-leading-order (NLO) approximation of
perturbative
QCD has been observed for $Q^2 \geq 2$ GeV$^2$ (see reviews in \cite{CoDeRo}
and references therein) and, thus,
perturbative QCD can describe the
evolution of $F_2$ and its derivatives
down to very low $Q^2$ values.

The standard program to study the $x$ behaviour of
quarks and gluons
is carried out comparing the experimental data
with the numerical solution of the DGLAP
equations 
\cite{DGLAP}
by fitting
the QCD energy scale $\Lambda$ and the parameters of the
$x$-profile of partons at some initial $Q_0^2$ \cite{fits,Ourfits}.
However, to investigate exclusively the
small-$x$ region, there is the alternative of doing the simpler analysis
by using some of the existing analytical solutions of DGLAP 
in the small-$x$
limit \cite{BF1}-\cite{HT}.
It was pointed out in \cite{BF1} that the HERA small-$x$ data can be
well interpreted in 
terms of the so-called doubled asymptotic scaling (DAS) phenomenon
related to the asymptotic 
behaviour of the DGLAP evolution 
discovered many years ago \cite{Rujula}.

The study of \cite{BF1} was extended in \cite{Munich}-\cite{HT}
to include the finite parts of anomalous dimensions (ADs)
of Wilson operators and Wilson coefficients\footnote{ 
In the standard DAS approximation \cite{Rujula} only the AD singular
parts 
were used.}.
This has led to predictions \cite{Q2evo,HT} of the small-$x$ asymptotic 
form of parton distribution functions (PDFs)
in the framework of the DGLAP dynamics,
which were obtained
starting at some $Q^2_0$ with
the flat function
 \begin{eqnarray}
f_a (Q^2_0) ~=~
A_a ~~~~(\mbox{hereafter } a=q,g), \label{1}
 \end{eqnarray}
where $f_a$ are PDFs
multiplied by $x$
and $A_a$ are unknown parameters to be determined from the data.

We refer to the approach of \cite{Munich}-\cite{HT}
as {\it generalized} DAS approximation. In this approach
the flat initial conditions, Eq. (\ref{1}), determine the
basic role of the AD singular parts 
as in the standard DAS case, while
the contribution from AD finite parts 
and from Wilson coefficients can be
considered as corrections which are, however, important for better 
agreement with experimental data. 

The use of the flat initial condition, given in Eq. (\ref{1}), is
supported by the actual experimental situation: low-$Q^2$ data
\cite{NMC,lowQ2,DIS02} are well described for $Q^2 \leq 0.4$ GeV$^2$
by Regge theory with Pomeron intercept
$\alpha_P(0) \equiv \lambda_P +1 =1.08$,
closed to the adopted ($\alpha_P(0) =1$) one. 
The small rise of HERA data \cite{H1ZEUS,:2009wt,lowQ2,lowQ2N}
at low $Q^2$ can be 
explained, for example, by contributions of
higher twist operators
(see \cite{HT}).

The purpose of this paper is to demonstrate a good agreement \cite{Cvetic1}
between 
the predictions of the generalized DAS approach \cite{Q2evo}
and the HERA experimental data
\cite{H1ZEUS} (see Fig. 1 below) for the structure function
(SF) $F_2$. We also compare the result
of
the slope $\partial  \ln F_2/\partial \ln (1/x)$ calculation
with the 
H1 and ZEUS 
data \cite{H1slo,DIS02}.
Looking at the H1 data \cite{H1slo} points shown in Fig. 2 
one can conclude 
that $\lambda (Q^2)$
is independent on $x$ within the experimental uncertainties
for fixed $Q^2$ in the range $x <0.01$. 
The
rise of $\lambda (Q^2)$ linearly with $\ln Q^2$ could be treated in strong 
nonperturbative way (see \cite{Schrempp} and references therein), i.e.,
$\lambda (Q^2) \sim 1/\alpha_s(Q^2)$. The 
analysis \cite{KoPa02}, 
however, demonstrated
that this rise can be explained naturally in the framework of 
perturbative QCD.

The ZEUS and H1 Collaborations have also presented \cite{DIS02} 
the preliminary data
for $\lambda (Q^2)$ at quite low values of $Q^2$.
The ZEUS value for 
$\lambda (Q^2)$ is consistent with a constant $\sim 0.1$ at $Q^2 <
0.6$ GeV$^2$, as it is expected under the assumption of single soft Pomeron
exchange within the framework of Regge phenomenology. 
It was important to extend the analysis of \cite{KoPa02} to low $Q^2$ range
with a help of well-known infrared modifications of the strong coupling 
constant. We used the ``frozen'' and analytic versions (see,  
\cite{Cvetic1}).

\section{
Generalized DAS
approach} \indent

The flat initial condition (\ref{1}) corresponds to the case when PDFs
tend  to some constant value at $x \to 0$ and at some initial value $Q^2_0$.
The main ingredients of the results \cite{Q2evo,HT}, are:
\begin{itemize}
\item
Both, the gluon and quark singlet densities
\footnote{The contribution of valence quarks is negligible at low $x$.}
 are presented in terms of two
components ($"+"$ and $"-"$) which are obtained from the analytic 
$Q^2$-dependent expressions of the corresponding ($"+"$ and $"-"$) PDF
moments.
\item
The twist-two part of the $"-"$ component is constant at small $x$ at any 
values of $Q^2$,
whereas the one of the $"+"$ component grows at $Q^2 \geq Q^2_0$ as
\begin{equation}
\sim e^{\sigma},~~~
\sigma = 2\sqrt{\left[ \hat{d}_+ s
- \left( \hat{D}_+ +  \hat{d}_+ \frac{\beta_1}{\beta_0} \right) p
\right] \ln \left( \frac{1}{x} \right)}  \ ,~~~ \rho=\frac{\sigma}{2\ln(1/x)} \ ,
\label{intro:1}
\end{equation}
where $\sigma$ and $\rho$
are the generalized Ball--Forte
variables,
\begin{equation}
s=\ln \left( \frac{a_s(Q^2_0)}{a_s(Q^2)} \right),~~
p= a_s(Q^2_0) - a_s(Q^2),~~~
\hat{d}_+ = \frac{12}{\beta_0},~~~
\hat{D}_+ =  \frac{412}{27\beta_0}.
\label{intro:1a}
\end{equation}
\end{itemize}
Hereafter we use the notation
$a_s=\alpha_s/(4\pi)$.
The first two coefficients of the QCD $\beta$-function in the $\ms$-scheme
are $\beta_0 = 11 -(2/3) f$
and $\beta_1 =  102 -(114/9) f$
with $f$ is being the number of active quark flavours.

Note here that the perturbative coupling constant $a_s(Q^2)$ is different at
the leading-order (LO) and NLO approximations. 
Hereafter we consider for simplicity only  the LO
approximation\footnote{
The NLO results may be found in  \cite{Q2evo}.},
where the variables 
$\sigma$ and $\rho$ are
given by Eq. (\ref{intro:1}) when $p=0$.

\subsection{Parton distributions and the structure function $F_2$
} 

The SF
$F_2$ and PDFs have the following form
\begin{eqnarray}
	F_2(x,Q^2) &=& e \, f_q(x,Q^2),~~
	f_a(x,Q^2) 
~=~ f_a^{+}(x,Q^2) + f_a^{-}(x,Q^2),~~(a=q,g)
\label{8a}
\end{eqnarray}
where
$e=(\sum_1^f e_i^2)/f$ is the average charge square.

The small-$x$ asymptotic results for PDFs
$f^{\pm}_a$ are
\begin{eqnarray}
	f^{+}_g(x,Q^2) &=& \biggl(A_g + \frac{4}{9} A_q \biggl)
		\tilde{I}_0(\sigma) \; e^{-\overline d_{+}(1) s} + O(\rho),~~
	f^{+}_q(x,Q^2) ~=~
\frac{f}{9} \frac{\rho \tilde{I}_1(\sigma)}{\tilde{I}_0(\sigma)}
+ O(\rho),
\nonumber \\
	f^{-}_g(x,Q^2) &=& -\frac{4}{9} A_q e^{- d_{-}(1) s} \, + \, O(x),~~
	f^{-}_q(x,Q^2) 
~=~ A_q e^{-d_{-}(1) s} \, + \, O(x),
	\label{8.02}
\end{eqnarray}
where $d_{-}(1) = 16f/(27\beta_0)$ and
$\overline d_{+}(1) = 1 + 20f/(27\beta_0)$ is
the regular part of AD
$d_{+}(n)$ 
in the limit $n\to1$\footnote{
We denote the singular and regular parts of a given quantity $k(n)$ in the
limit $n\to1$ by $\hat k(n)$ and $\overline k(n)$, respectively.}.
Here $n$ is
the variable in Mellin space.
%
%
The functions $\tilde I_{\nu}$ ($\nu=0,1$) 
are related to the modified Bessel
function $I_{\nu}$
and to the Bessel function $J_{\nu}$ by:
\begin{equation}
\tilde I_{\nu}(\sigma) =
\left\{
\begin{array}{ll}
I_{\nu}(\sigma), & \mbox{ if } s \geq 0 \\
i^{-\nu} J_{\nu}(i\sigma), \ i^2=-1, \ & \mbox{ if } s \leq 0 
\end{array}
\right. .
\label{4}
\end{equation}

\subsection{Effective slopes}

As it has been 
shown in \cite{Q2evo},
the behaviour of PDFs
and $F_2$ given in the Bessel-like form 
by generalized DAS approach
can mimic a power law shape
over a limited region of $x$ and $Q^2$
 \begin{eqnarray}
f_a(x,Q^2) \sim x^{-\lambda^{\rm eff}_a(x,Q^2)}
 ~\mbox{ and }~
F_2(x,Q^2) \sim x^{-\lambda^{\rm eff}_{\rm F_2}(x,Q^2)}.
\nonumber    \end{eqnarray}

The effective slopes 
$\lambda^{\rm eff}_a(x,Q^2)$ and $\lambda^{\rm eff}_{\rm F_2}(x,Q^2)$
have the form:
 \begin{eqnarray}
\lambda^{\rm eff}_{\rm F_2}(x,Q^2) &=& 
\lambda^{\rm eff}_g(x,Q^2) ~=~
\frac{f^+_g(x,Q^2)}{f_g(x,Q^2)} \,
\rho \, \frac{\tilde I_1(\sigma)}{\tilde I_0(\sigma)}\approx \rho - 
\frac{1}{4\ln{(1/x)}},
\nonumber
\\
\lambda^{\rm eff}_q(x,Q^2) &=& \frac{f^+_q(x,Q^2)}{f_q(x,Q^2)} \,
\rho \, \frac{\tilde I_2(\sigma)}{\tilde I_1(\sigma)} \approx 
 \rho - \frac{3}{4\ln{(1/x)}} 
,
\label{10.1}
\end{eqnarray}
where 
the symbol $\approx $ marks the approximation obtained in the  expansion
of the modified Bessel functions, when 
the ``$-$'' component is negligible.
These approximations are
accurate only at very large $\sigma $ values (i.e. at very large $Q^2$
and/or very small $x$).

\begin{figure}[t]
\centering
\vskip-0.5cm
\includegraphics[width=.75\hsize]{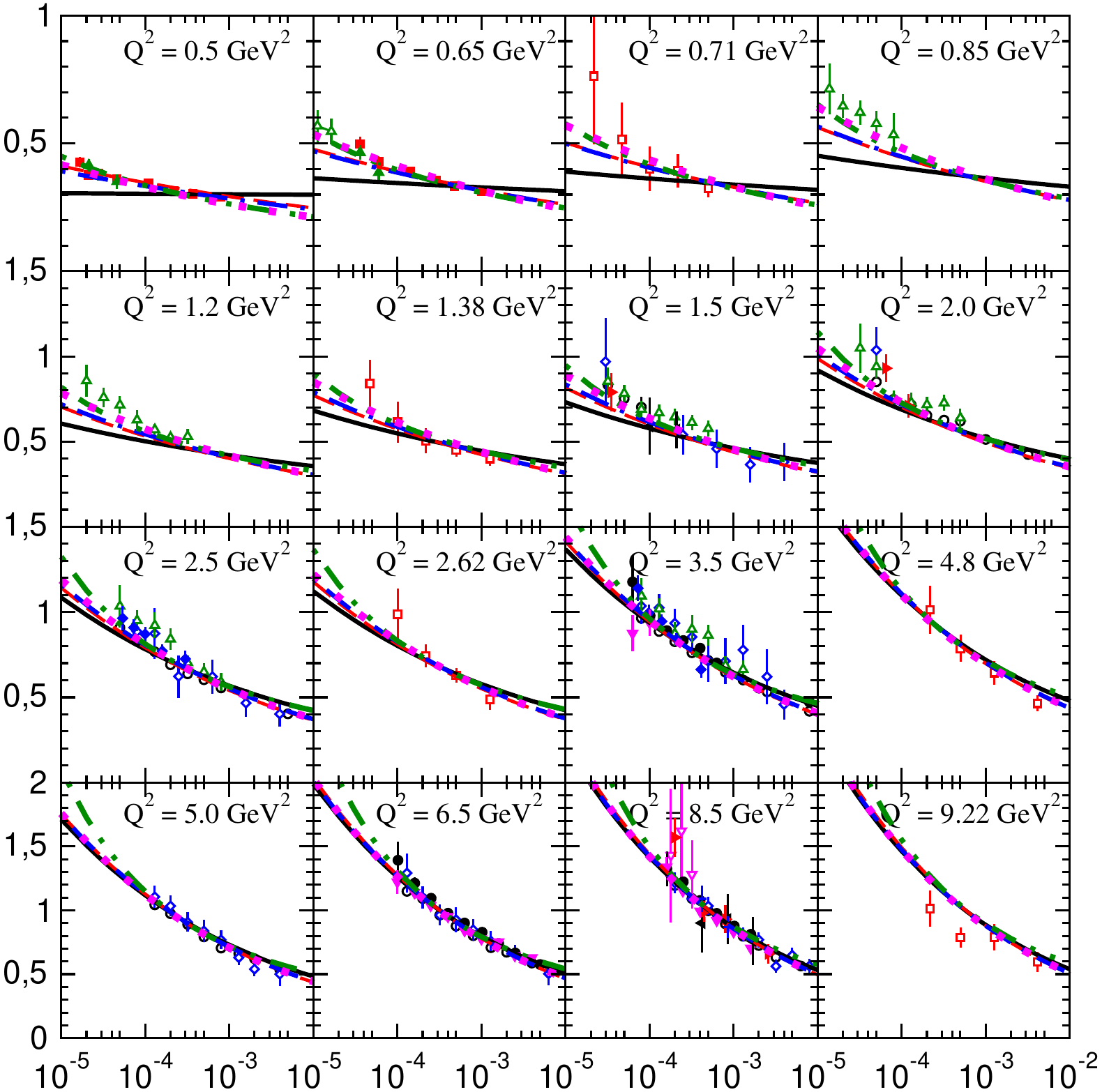}
\vskip -0.3cm
\caption{$x$ dependence of $F_2(x,Q^2)$ in bins of $Q^2$.
The experimental data from H1 (open points) and ZEUS (solid points) 
\cite{H1ZEUS} are
compared with the NLO fits for $Q^2\geq0.5$~GeV$^2$ implemented with the
canonical (solid lines), frozen (dot-dashed lines), and analytic (dashed lines)
versions of the strong-coupling constant.
}
\label{fig:F2}
\end{figure}

\section{Comparison with experimental data
} \indent

Using the results of previous section we have
analyzed  HERA data for $F_2$ \cite{H1ZEUS} and 
the slope $\partial \ln F_2/\partial \ln (1/x)$ \cite{H1slo,DIS02}
at small $x$ from the H1 and ZEUS Collaborations.
In order to keep the analysis as simple as possible,
we fix $f=4$ and $\alpha_s(M^2_Z)=0.1166 $ (i.e., $\Lambda^{(4)} = 284$ MeV) 
in agreement with the recent ZEUS results in \cite{H1ZEUS}.

As it is possible to see in Figs. 1 and 2, 
the twist-two
approximation is reasonable at $Q^2 \geq 2\div 4$ GeV$^2$. 
At smaller $Q^2$,  some
modification of the approximation should be considered. In Ref.
\cite{HT} we have added the higher twist corrections.
For renormalon model of higher twists, we
have found a good
agreement with experimental data at essentially lower $Q^2$ values:
$Q^2 \geq 0.5$ GeV$^2$ (see Figs. 4 and 5 in \cite{HT}).

In Ref. \cite{Cvetic1},
to improve the agreement at small $Q^2$ values,
we modified the QCD coupling constant.
We consider two modifications. 

In one case, which is more phenomenological, we introduce freezing
of the coupling constant by changing its argument $Q^2 \to Q^2 + M^2_{\rho}$,
where $M_{\rho}$ is the $\rho $-meson mass (see \cite{Cvetic1} and 
references therein). Thus, in the 
formulae of the
Section 2 we should do the following replacement:
\begin{equation}
 a_s(Q^2) \to a_{\rm fr}(Q^2) \equiv a_s(Q^2 + M^2_{\rho})
\label{Intro:2}
\end{equation}

\begin{figure}[t]
\centering
\vskip-0.5cm
\includegraphics[width=.8\hsize]{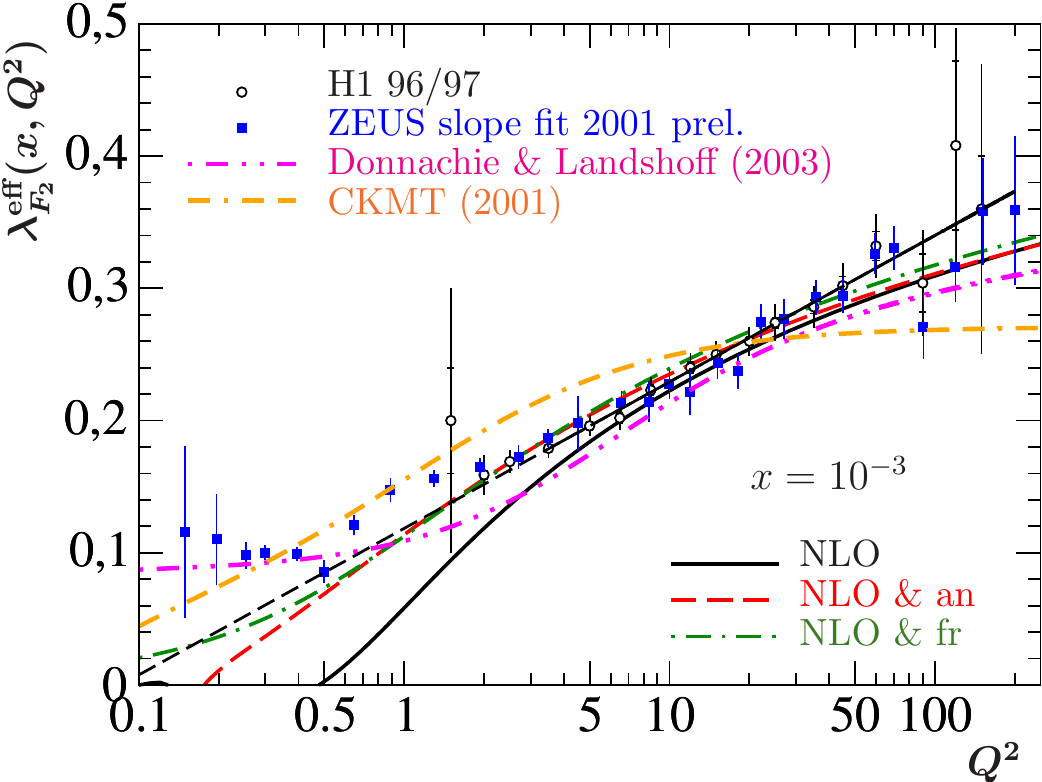}
\vskip -0.3cm
\caption{
As in Fig,1 but for the $Q^2$ dependence of 
$\lambda^{\rm eff}_{F_2}(x,Q^2)$ for an average
small-$x$ value of $x=10^{-3}$.
The linear rise of $\lambda^{\rm eff}_{F_2}(x,Q^2)$ with $\ln Q^2$ 
\cite{H1slo}
is indicated by the straight dashed line.
For comparison, also the results obtained in the phenomenological models
by Kaidalov et al. \cite{Kaidalov} (dash-dash-dotted line)
and by Donnachie and Landshoff \cite{Donnachie:2003cs} (dot-dot-dashed 
line) are shown.
}
\label{fig:Q2-slope_lin}
\end{figure}

\begin{figure}[t]
\centering
\vskip-0.5cm
\includegraphics[width=.8\hsize]{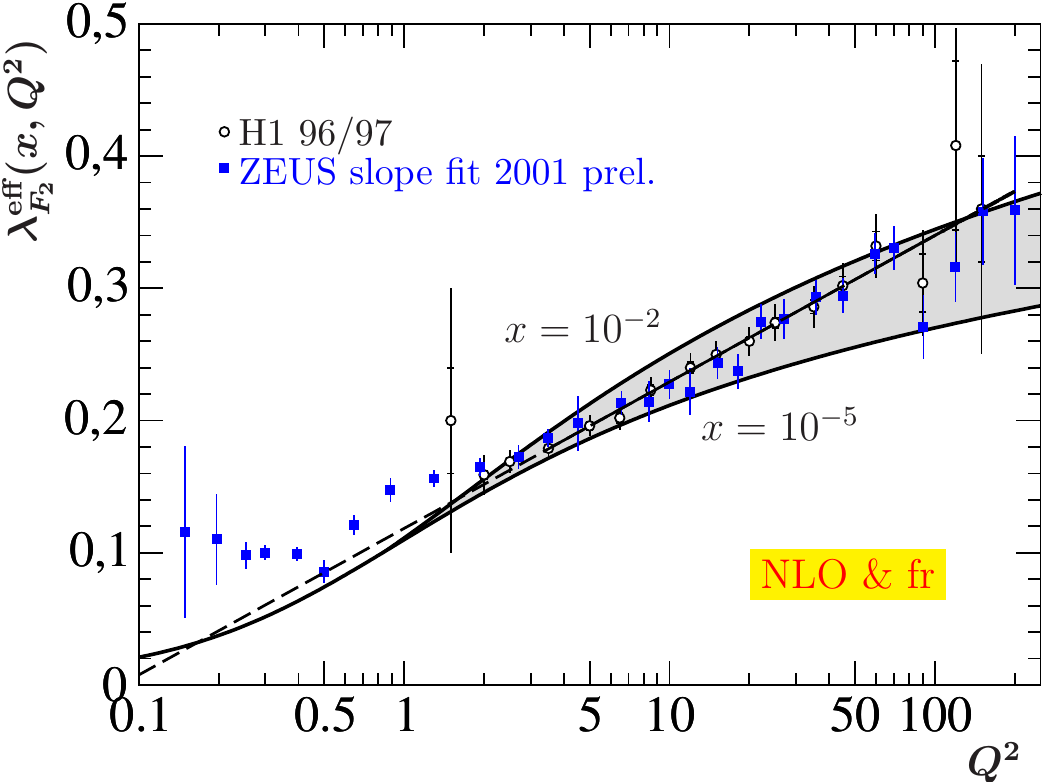}
\vskip -0.3cm
\caption{
The values of effective slope
$\lambda^{\rm eff}_{\rm F_2}$  
as a function of $Q^2$.
The experimental points are same as on Fig. 4.
The dashed line
 represents the fit from \cite{H1slo}.
The solid curves represent the NLO fits with 
``frozen'' coupling constant
at $x=10^{-2}$ and  $x=10^{-5}$.}
\label{fig:Q2-slope}
\end{figure}

The second possibility incorporates the Shirkov--Solovtsov idea 
\cite{ShiSo}
about analyticity of the coupling constant that leads to the additional its
power dependence. Then, in the formulae of the previous section
the coupling constant $a_s(Q^2)$ should be replaced as follows:
($k=1$ and $2$ at LO and NLO)
\begin{eqnarray}
 a_{\rm an}(Q^2) \, = \, a_s(Q^2) - \frac{1}{k\beta_0}
 \frac{\Lambda^2}{Q^2 - \Lambda^2} 
+ \ldots \, ,
\label{an:NLO}
\end{eqnarray}
where the symbol $\ldots$ stands for terms which are zero and 
negligible at $Q \geq 1$ GeV \cite{ShiSo} at LO and NLO, respectively.


Figure~2 shows the experimental data for $\lambda_{F_2}^{\rm eff}(x,Q^2)$
at $x\sim 10^{-3}$, which represents an average of the $x$-values of 
HERA experimental 
data. The top dashed line represents the aforementioned linear rise of
$\lambda(Q^2)$ with $\ln(Q^2)$.
The Figs. 1 and 2
demonstrate
that the theoretical description of the small-$Q^2$ ZEUS
data for $\lambda^{\rm eff}_{F_2}(x,Q^2)$ by NLO QCD is significantly
improved by implementing the ``frozen'' and analytic coupling constants
$\alpha_{\rm fr}(Q^2)$ and $\alpha_{\rm an}(Q^2)$, 
respectively,
which in turn lead to
very close results (see also \cite{KoLiZo}).

Indeed, the fits for $F_2(x,Q^2)$ in \cite{HT} yielded
$Q^2_0 \approx 0.5$--$0.8$~GeV$^2$.
So, initially we had $\lambda^{\rm eff}_{F_2}(x,Q^2_0)=0$,
as suggested by Eq.~(\ref{1}). The replacements of Eqs.~(\ref{Intro:2})
and (\ref{an:NLO}) modify the
value of $\lambda^{\rm eff}_{F_2}(x,Q^2_0)$. 
For the  
``frozen'' and analytic coupling constants 
$\alpha_{\rm fr}(Q^2)$ and $\alpha_{\rm an}(Q^2)$,
the value of
$\lambda^{\rm eff}_{F_2}(x,Q^2_0)$ is non-zero 
and the slopes are
quite close to the experimental data at $Q^2 \approx 0.5$~GeV$^2$.
Nevertheless, for $Q^2 \leq 0.5$~GeV$^2$, there is still some disagreement with
the data, which needs additional investigation.

For comparison, we display in Fig. 2 also the results obtained by Kaidalov 
et al. \cite{Kaidalov}
and by Donnachie and Landshoff \cite{Donnachie:2003cs} 
adopting phenomenological models based on Regge theory. While they yield 
an improved description of the experimental data for $Q^2\leq 0.4$ GeV$^2$, 
the agreement generally worsens in the range $2$ GeV$^2 \leq Q^2 \leq 8$ 
GeV$^2$.

 The results of fits in \cite{HT,Cvetic1} have an important property: they are
very similar in LO and NLO approximations of perturbation theory.
The similarity is related to the fact that the small-$x$ asymptotics of 
the NLO corrections
are usually large and negative 
(see, for example, $\alpha_s$-corrections \cite{FaLi} to
BFKL approach
\cite{BFKL}
\footnote{It seems that it is a property of 
any processes in which gluons,
but not quarks play a basic role.}).
Then, the LO form $\sim \alpha_s(Q^2)$ for
some observable and the NLO one 
$\sim \alpha_s(Q^2) (1-K\alpha_s(Q^2)) $
with a large value of $K$, are similar because 
$\Lambda \gg
\Lambda_{\rm LO}$\footnote{The equality of
$\alpha_s(M_Z^2)$ at LO and NLO approximations,
where $M_Z$ is the $Z$-boson mass, relates $\Lambda$ and $\Lambda_{\rm LO}$:
$\Lambda^{(4)} = 284$ MeV (as in ZEUS paper on \cite{H1ZEUS}) corresponds to 
$\Lambda_{\rm LO} = 112$ MeV (see \cite{HT}).}
and, thus, $\alpha_s(Q^2)$ at LO is considerably smaller  then 
$\alpha_s(Q^2)$ at NLO  for HERA $Q^2$ values.

In other words, performing some resummation procedure (such as Grunberg's 
effective-charge method \cite{Grunberg}), one can see that the 
NLO form may
be represented as $\sim \alpha_s(Q^2_{\rm eff})$,
where $Q^2_{\rm eff} \gg Q^2$. 
Indeed, from 
different studies
\cite{DoShi},
it is well known that at small-$x$ values the effective
argument of the coupling constant is higher then $Q^2$.

In the generalized DAS approach the small effect
of the NLO corrections
can be explained by separated contributions of the singular and regular
AD parts. Indeed, the singular parts modify the argument of the Bessel 
functions (see Eq.(\ref{intro:1})) and the regular parts contribute to the
front of Bessel functions \cite{Q2evo}.

Figure~3 shows the $x$-dependence of the slope 
$\lambda^{\rm eff}_{F_2}(x,Q^2)$.
One observes good agreement between the experimental data and the generalized
DAS approach for a broad range of small-$x$ values.
The absence of a variation with $x$ of 
$\lambda^{\rm eff}_{F_2}(x,Q^2)$ at small $Q^2$ values is related to the small
values of the variable $\rho$ there.

From  Figs. 2 and 6 in \cite{HT}, 
one can see that HERA experimental data exists
at $x \sim 10^{-4} \div 10^{-5}$
for $Q^2=4$~GeV$^2$ and at $x \sim 10^{-2}$ for $Q^2=100$~GeV$^2$. Indeed,
the correlations between $x$ and $Q^2$  in the form 
$x_{\rm eff}= a \times 10^{-4} \times Q^2$ with $a=0.1$ and $1$
lead
to a modification of the $Q^2$ evolution which starts
to resemble $\ln Q^2$, rather than $\ln \ln Q^2$ as is standard 
\cite{KoPa02}.

\section{Conclusions} \indent

We have shown
the $Q^2$-dependence of the SF
$F_2$ and the slope 
$\lambda^{\rm eff}_{F_2}=\partial \ln F_2/\partial \ln (1/x)$ at 
small-$x$ values in the 
framework of perturbative QCD. Our twist-two 
results are in very good agreement with 
the precise HERA data at $Q^2 \geq 2$~GeV$^2$,
where the perturbative theory is applicable.
The application of the ``frozen'' and analytic coupling constants 
$\alpha_{\rm fr}(Q^2)$
and $\alpha_{\rm an}(Q^2)$ improves
the agreement 
for smaller $Q^2$ values, down to
$Q^2 \geq 0.5$~GeV$^2$.

As a next step of investigations, we plan to fit the H1$\&$ZEUS data 
\cite{:2009wt} and to extend the generalized DAS approach to evaluate
the double PDFs which are very popular now (see \cite{Snigirev} and 
references therein). Also we plan to use our approach to analyse the
cross sections of processes studied at LHC by analogy with our
investigations \cite{Fiore}
of the total cross section of ultrahigh-energy deep-inelastic  
neutrino-nucleon scattering. \\

\vspace{-0.3cm}
 
A.V.K. thanks the Organizing Committee of 
the 3rd International Workshop on Multiple Partonic
Interactions at the LHC
for invitation and support. This work was
supported in part by RFBR grant 11-02-01454-a.


\end{document}